\documentclass{article}

    \PassOptionsToPackage{numbers, compress}{natbib}

\usepackage[final]{neurips_2021}

\setcitestyle{square}
\usepackage[utf8]{inputenc} 
\usepackage[T1]{fontenc}    
\usepackage{url}            
\usepackage{booktabs}       
\usepackage{amsfonts}       
\usepackage{subfigure}
\usepackage{pdfpages}
\usepackage{amsmath}
\usepackage[colorinlistoftodos]{todonotes}
\usepackage{nicefrac}       
\usepackage{microtype}      
\usepackage{xcolor}         
\usepackage{url}

\usepackage{breakurl}
\usepackage[breaklinks]{hyperref}
\usepackage[inline]{enumitem}
\usepackage{graphics}
\usepackage{multirow}
\usepackage{tabularx,ragged2e,booktabs}

\title{Weight, Block or Unit? Exploring Sparsity Tradeoffs for Speech Enhancement on Tiny Neural Accelerators}

\author{
  Marko Stamenovic \\
  Bose Corp. \\
  Boston, MA \\
  \texttt{marko\_stamenovic@bose.com} \\
  \And
  Nils L. Westhausen \\
  University of Oldenburg and Bose Corp.  \\
  Oldenburg, Germany \\
  \texttt{nils.westhausen@uol.de} \\
  \AND
  Li-Chia Yang \\
  Bose Corp.  \\
  Boston, MA \\
  \And
  Carl Jensen \\
  Bose Corp.  \\
  Boston, MA \\
  \And
  Alexander Pawlicki \\
  Bose Corp.  \\
  Boston, MA \\
}

\begin{document}

\maketitle
 
\begin{abstract}
We explore network sparsification strategies with the aim of compressing neural speech enhancement (SE) down to an optimal configuration for a new generation of low power microcontroller based neural accelerators (microNPU's). We examine three unique sparsity structures: weight pruning, block pruning and unit pruning; and discuss their benefits and drawbacks when applied to SE. We focus on the interplay between computational throughput, memory footprint and model quality. Our method supports all three structures above and jointly learns integer quantized weights along with sparsity. Additionally, we demonstrate offline magnitude based pruning of integer quantized models as a performance baseline. Although efficient speech enhancement is an active area of research, our work is the first to apply block pruning to SE and the first to address SE model compression in the context of microNPU's. Using weight pruning, we show that we are able to compress an already compact model's memory footprint by a factor of 42$\times$ from 3.7MB to 87kB while only losing 0.1 dB SDR in performance. We also show a computational speedup of 6.7 $\times$ with a corresponding SDR drop of only 0.59 dB SDR using block pruning.
\end{abstract}

\section{Introduction}
 Recent years have seen a step function increase in performance of deep learning based SE and source separation in the research community \citep{trends}. Consequently, SE has recently gained wide adoption in industry and has quickly become a key piece of the modern telecommunications stack with applications from voice communication \cite{teams, meet, pinpoint} to audio compression \citep{lyra, dolby, soundstream} to studio production \citep{izotope, descript}. In a short time, these applications have redefined what users have come to expect from their audio devices. However, the models that power modern SE applications can often be large and processor-intensive. Computation is generally performed on powerful mobile, desktop or cloud hardware with access to robust processing and abundant power. 
 
 While there has been continued interest in real-time augmented hearing applications on battery powered wearable devices, from hearing aids to consumer products\citep{hearphones, nuheara, airpods}, most of these experiences are currently powered by non AI-based signal processing. SE has long shown promise for hearing aid (HA) applications in particular \citep{healy}, but its successful implementation has been elusive due to the 
 constraints imposed by the HA hardware and use-case.

 Historically, the SE literature has prioritized objective and perceptual audio quality \citep{chimera++, dpcl++, furcanext, upitblstm, tasnet, wavesplit} over model efficiency or realizability.
 Recently, however, there has been a flurry of research with the goal of high quality SE which can also be deployed in realtime use-cases and run on streaming audio. In \citep{tinylstms}, Fedorov \textit{et. al.} design a SE framework to satisfy hearing aid hardware constraints, restricting their attention to unit pruning. Choi \textit{et. al.} target similar hardware in \citep{trunet} where they describe a parameter efficient network which nevertheless consumes approximately 2.27 billion operations per second (GOPS/s), far beyond the computational limits of modern MCU's and microNPU's. Tan and Wang perform weight and unit pruning across various SE architectures along with Huffman Coding based quantization \citep{delian} using an iterative offline approach. Other studies focus on latency and computational complexity of full precision SE networks \citep{dtln, sudormrf}. 

Concurrently, a new generation of micro-controller-based neural processing units (microNPU's) have been surfacing \citep{tensilica, u55, edgeinf}, promising low-power embedded execution of AI workloads such as efficient SE. There are some similarities and trends we identify across the emerging landscape of edge neural network accelerators including 1) a fixed point multiply-accumulate (MAC) unit which is able to parallelize small amounts of vector-vector operations over a single CPU cycle and 2) the ability to support sparse tensor data types and operations to further reduce memory footprint, increase efficiency and potentially speed execution.

 We enumerate four specific constraints which must be met in order for a neural SE algorithm to be deployable on a microcontroller-based hearable device based on Fedorov \textit{et. al}. \cite{tinylstms}: 
 \begin{enumerate}

 \item The model must be causal in order to enable inference on streaming audio. Total audio latency must be kept at a minimum, ideally below 30ms.
 \item Model inputs, weights and activations must be quantized to symmetrical fixed point precision such that they can run on embedded hardware processors which often do not support floating point execution
 \item Computational latency, or the time it takes to compute one inference of the model, must be less than or equal to the time until the next inference is required. For frame-based signal processing this also known as the frame or hop rate. 
 \item Total memory footprint, the sum of the model weights and working memory (model activations), must fit within the device's SRAM. 
  \end{enumerate}

 Our goal is to study the effects of compression on SE in the context of real-time edge deployment on microNPU’s. We present various methodologies to compress and quantize SE networks in order to meet the constraints defined above with a focus on trade-offs in computational throughput, memory footprint and model quality.
Firstly, we demonstrate an offline (post-training) method for pruning SE models in both unstructured and structured regimes. 
Next, we extend the online pruning method from \citep{tinylstms} to support both unstructured weight and structural block pruning.
Although block pruning has been studied elsewhere in the literature \citep{blocksparse, blocksparsernn}, it is the first time to our knowledge that this type of structural sparsity has been applied to speech enhancement.
We further propose a simple and efficient method for symmetrically quantizing networks for SE which we refer to as Quantization EQ. We use this method alongside all of our sparsification strategies. 
Finally, we compare the online method for inducing sparsity to the offline method and provide benchmarks for the three types of sparsity that we explore, namely unit block and weight, on the popular CHiME2 \citep{chime2} dataset.








\section{Background}

\subsection{Speech Enhancement}

Given a noisy time-domain monaural audio mixture $y = x + v$ where $x$ is the clean speech signal and $v$ is unwanted noise, the goal of SE is to extract the denoised speech signal $\hat{x}$. We formulate neural SE as 
\vspace{-6pt}
\begin{align} \label{eq:se}
Y &= \mathcal{F}(y),    &     Z &= (G |Y|)^{0.3}, \\
M &= f_\theta(Z),       &    \hat{x} &= \mathcal{I}(G^TM \odot Y) \nonumber
\end{align}
\vspace{-6pt}

where $\mathcal{F}$ and $\mathcal{I}$ denote forward and inverse short time Fourier transforms (STFT), respectively. The output of $\mathcal{F}$ is a complex STFT $Y \in \mathbb{C}^{B_f \times B_t}$ where $B_f$ is number of frequency bins and $B_t$ is number of STFT frames. $Y$ is then preprocessed by taking the magnitude and passing through a mel-filterbank $G \in \mathbb{R}^{B_g \times B_f}$ where $B_g$ is the mel dimension, before being power law compressed with an exponent of 0.3. The preprocessing step is done to map to a logarithmic frequency scale, compress dynamic range,  and remove phase, respecively. Our SE network $f$, parametrized by learnable parameters $\theta$, predicts a time-frequency mask in the mel space $M \in \mathbb{R}^{B_g \times B_t}$. We invert $M$ to the frequency domain using the transpose $G^T$ and apply the mask to $Y$ using pointwise multiplication $\odot$ before resynthesizing with $\mathcal{I}$, resulting in the denoised time-domain speech signal $\hat{x}$. 

The SE network parameters, $f_\theta(\cdot)$, are learned by minimizing a phase-sensitive spectral approximation loss \cite{erdogan2015phase}:
\vspace{-3pt}
\begin{equation}
    {L}(\theta) = 0.1\left \Vert \lvert X \rvert^{0.3} - \lvert \hat{X} \rvert^{0.3} \right \Vert_2 + 0.9  \left \Vert X^{0.3} - \hat{X}^{0.3} \right \Vert_2 \label{eq:psa_loss},
\end{equation}\vspace{-12pt}

where $X$ are the clean target frames and $\hat{X} = \mathcal{F}(\hat{x})$ are the output frames calculated from the output time series $\hat{x}$. The extra STFT transform is done to ensure consistency in the output frames \cite{diff_consistency}. The frames are also power-law compressed with an exponent of $0.3$ to reduce the dominance of large values.






\begin{figure}
    \centering
    \begin{minipage}[t]{0.48\textwidth}
        \centering
        \includegraphics[width=0.95\textwidth]{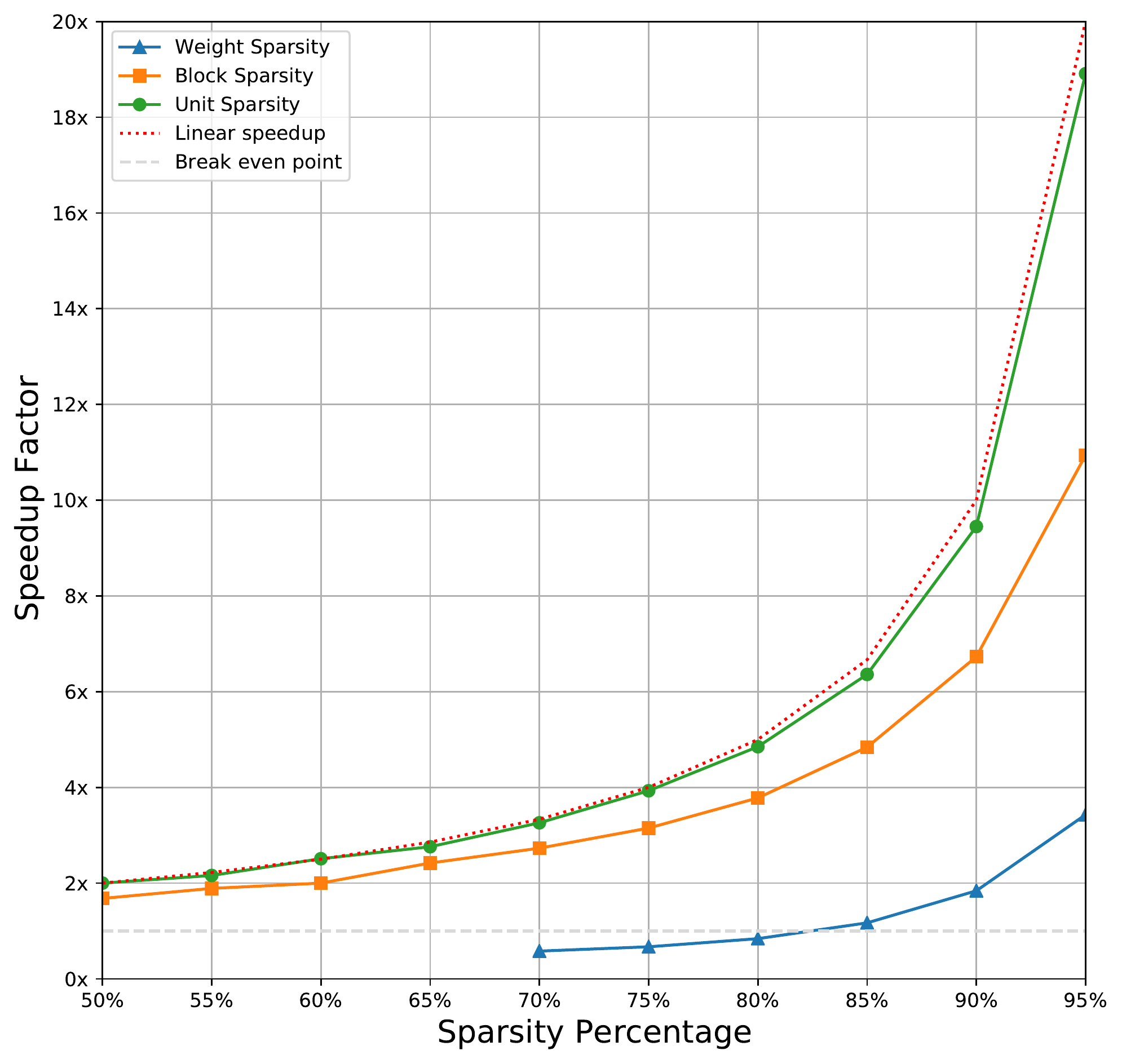}
        \caption{Plot of sparsity vs. computational speedup based on data aggregated from \citep{structuredsparsity, blocksparse} for the three sparsity structures we consider.}
         \label{fig:speedup}
    \end{minipage}\hfill
    \begin{minipage}[t]{0.48\textwidth}
        \centering
        \includegraphics[width=0.95\textwidth]{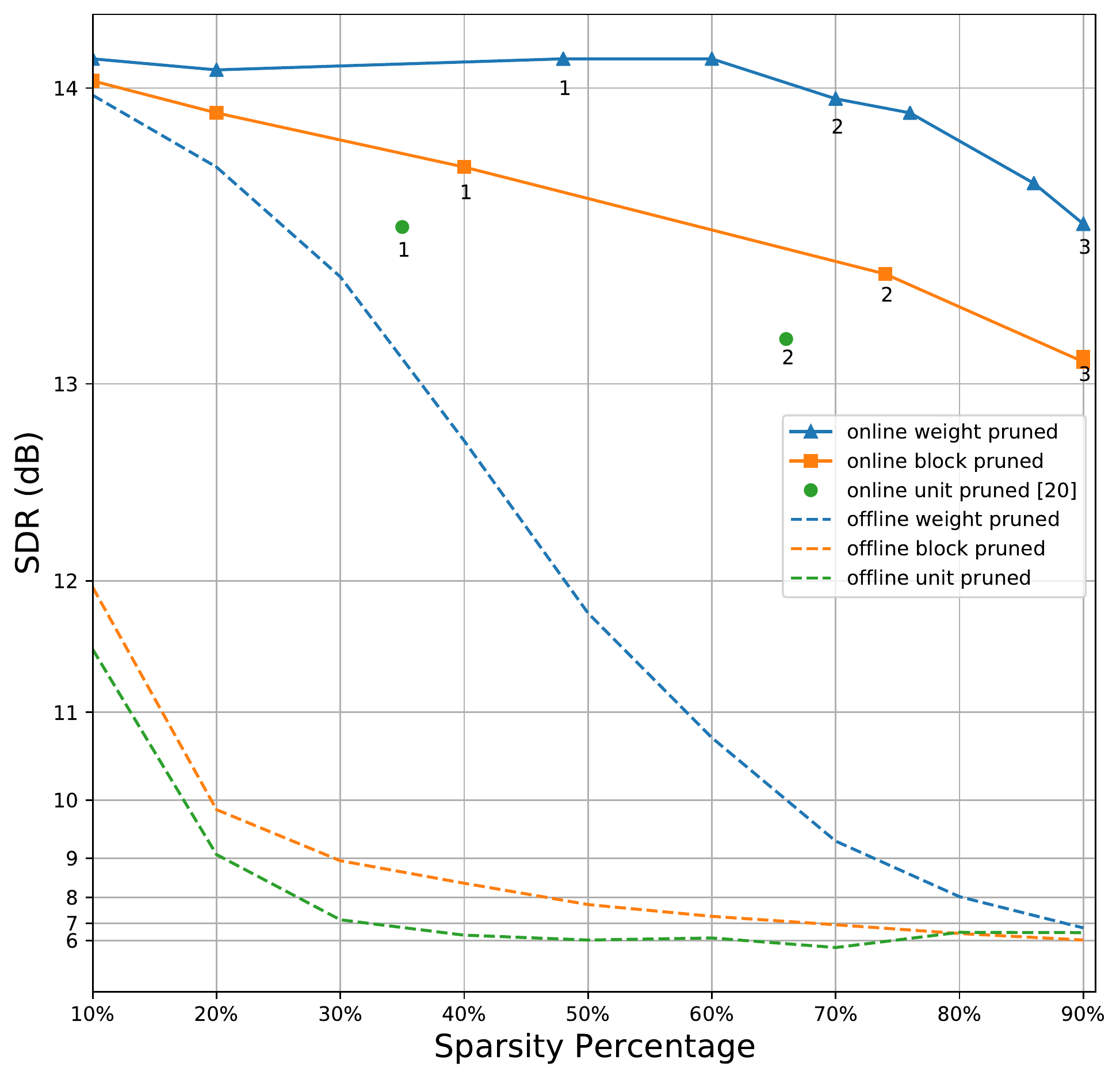} 
        \caption{SDR vs. Sparsity percentage, each point represents a model checkpoint presented in Table \ref{table:optimizations}. }
        \label{fig:sparsity}
    \end{minipage}
\end{figure}

\subsection{Baseline Model Architecture}

Our goal is to study the effects of compression on SE in the context of real-time edge deployment on microNPU's. We begin with a unidirectional recurrent architecture due to its causality and ability to effectively model time while retaining a frame-in, frame-out inference regime. This is in contrast to other architectures such as for example temporal convolutions \citep{tcn, wavenet} or transformers \citep{attention_is_all} which can yield impressive performance on time-series data but require an explicit, often large, receptive field which must be cached in memory to yield acceptable latency. We note that caching large amounts of dense audio can quickly overwhelm working memory constraints on embedded hardware.

Our network consists of two LSTM\citep{lstm} layers with hidden states of dimensionality 256 and have update rules $i^t = \sigma ( W_{xi} x^t + h^{t-1} W_{hi} + b_i ), f^t = \sigma ( W_{xf} x^t +  h^{t-1} W_{hf} + b_f ), o^t = \sigma ( W_{xo} x^t + h^{t-1} W_{ho} + b_o ), u^t = \tanh ( W_{xc} x^t + h^{t-1} W_{hc} + b_u ), c^t = f^t \odot c^{t-1} + i^t \odot u^t, h^t = o^t \odot \tanh \left( c^t \right)$. The LSTMs are followed by two dense layers of dimensionality 128; our first dense layer uses a $tanh$ nonlinearity to bound the dynamic range of our activations for quantization while the second one uses a $sigmoid$ in order to bound our mel mask prediction. We configure the network to operate on a frame size of 512 samples and hop size of 256 samples, with a 128 bin mel projection.
During training, we perform data augmentation by re-framing the speech and noise into 0.8s segments.
and continuously shuffling and mixing the speech with noise at an SNR generated by a uniform distribution determined by the SNR limits of the dataset (e.g. 
$U(-6, 9)$ dB for CHiME2). Note that we calculate SNR using Loudness units full scale (LUFS) \citep{lufs} in order to give a more perceptually meaningful SNR and avoid over-saturating mixtures where noise is dominated by small amounts of impulsive noise. Finally, we apply an additional amplitude scaling of $U(-5,5)$ dB on top of the noisy mixture.

\section{Model Compression}

Network connection pruning is a well-studied approach to compress neural  models \citep{optimalbraindamage, skeletonization, deepcompression}. Our contribution centers on the exploration of various compression schemes applied to SE under the constraints of battery-powered microNPU's.

\subsection{Quantization EQ}

We adopt a standard training-aware quantization approach based on Benoit et al. \citep{quantization}. The model inputs, weights, and activations are all quantized to 8 bits, while the final layer is quantized to 16 bits for improved output quality.

We design our model to use a [-1, 1] quantization dynamic range throughout the layers. This approach eliminates the need to learn dynamic ranges for each quantization node, but introduces the risk of information loss from overflow.
 To compensate, we propose a simple mechanism called Quantization EQ (QEQ) where a learnable gain vector, $g$, and bias, $b$, are applied to the input features from \eqref{eq:se}, changing them to $\hat{Z} = g \odot (G|Y|)^{0.3} + b$. This simple modification to the inputs is sufficient to maintain high performance from the fixed dynamic range used throughout the rest of the model.



\subsection{Sparse Structures}

We begin by grouping weights in $\theta$ into a set $\Gamma$, where $w_g \in \Gamma$ denotes the set of weights in a particular group. The organization of the groups defines the kind of structures we can learn and how they might interact with particular computer hardware. General sparse linear algebra operations offer speedups only for highly sparse matrices in most compute hardware, while other pruning structures might better reflect the underlying computer architecture and produce a more worthwhile trade-off between sparsity and performance.


This tendency can be seen in Fig. \ref{fig:speedup}, where sparsity is plotted vs. computational speedup for three types of pruning structures: weight, block and unit. Unit pruning achieves nearly linear speedup while block pruning shows slightly sublinear performance. Weight pruning shows negative speedup below approximately 85\% sparsity. These structures will be discussed in more detail below but the data demonstrates the significant differences in performance at a given level of sparsity for different types of pruning. 
The data in the figure is an aggregation from studies examining relative speedups of unit and weight pruning on CPUs \citep{structuredsparsity} and block pruning on GPUs \citep{blocksparse}. We adopt this as a guide for evaluating speedup of computational throughput in microNPUs across the rest of the paper. Visualization of the effects of each grouping structure are demonstrated in Fig. \ref{fig:weight}. 


\subsubsection{Weight Pruning}

Weight pruning operates on individual weights $w_g$ in matrices and not groups. As such it is considered a form of unstructured or random pruning, since the sparsity is dispersed randomly throughout the weight matrices. We apply weight pruning uniformly to both dense and LSTM layers.

\begin{figure}
  \centering
  \includegraphics[width=\columnwidth]{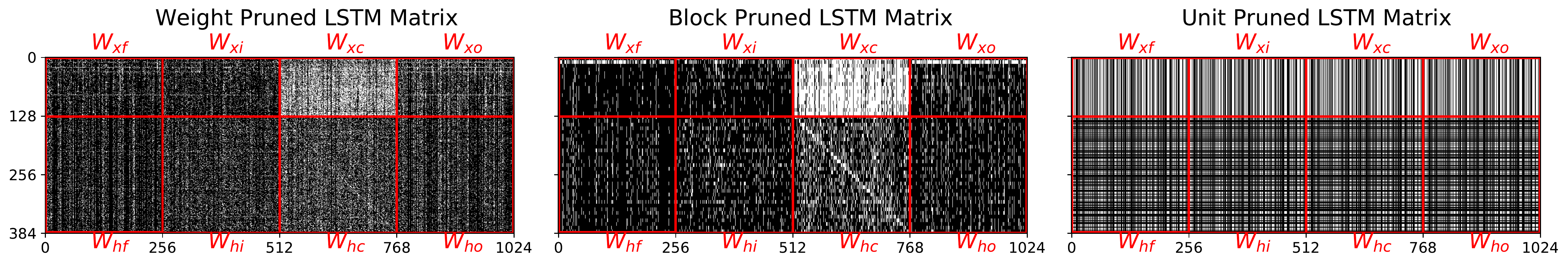}
  \caption{Visualizations of sparsity in a ~60\% sparse LSTM weight matrix in weight, block, and unit pruning. In weight and block pruning, we note that the $W_C$ matrices retain less sparsity than the gate matrices, indicating a higher importance to overall model performance. 
  Black pixels indicate weights that are pruned and white pixels indicate weights that remain.}
    \label{fig:weight}

\end{figure}

\subsubsection{Unit Pruning}

In unit pruning, a form of structural pruning, we aim to consider groups $\Gamma$ corresponding to entire rows of a weight matrix, or neurons. Numerous works have found that unit pruning leads to favorable computational throughput compared to unstructured pruning \citep{structuredsparsity, sparsityrnns, deepbench}. Additionally, unit pruning does not require any specialized hardware instruction-set for sparse data-types or operations, since unit pruning is equivalent to simply decreasing the size of the weight matrix, making it especially attractive for resource constrained microcontrollers. 

For dense layers we simply adopt a row-wise grouping of the weights with exception of the output layer, which we do not prune to preserve the output shape. For LSTM layers, however, we group weights based on the index of the LSTM output, $h^t$, to which they are connected, similar to \cite{structuredsparsity}. This grouping is necessary to avoid intra-LSTM shape inconsistencies that could arise from pruning different LSTM weight matrices at different rates.


\subsubsection{Block Pruning}

Block pruning is another form of structural pruning in which weights are grouped into blocks $\Gamma \in \mathbb{R}^{B_w \times B_h}$ where $B_w$ corresponds to the width of the block and $B_h$ the height. We note that unit pruning can be considered a special case of block pruning where ${B_w = B_r}$ and ${B_h = 1}$. In block pruning, both dense and LSTM layers are pruned in the same way, since we are not structurally affecting the shape of LSTM sub-matrices, in contrast to unit pruning.

Previous work \citep{blocksparse, blocksparsernn} has found that block sparsity can lead to large throughput gains on parallel processors. Additionally, since blocks are often more compact than rows (in our experiments, we consider $B_w=8 \times B_h=1$) they allow for more fine-grained pruning of the weight matrices, potentially resulting in higher model quality at equal sparsity levels to unit pruning.

Our motivation for investigating block pruning arises from the desire to capitalize on the parallelism of the new class of edge AI accelerators introduced in Section 1. We do so by inducing network sparsity in such a way as to skip entire computational cycles, a form of hardware-aware model design. Concretely, an edge accelerator such as the Tensilica HiFi 1 can process 4 16-bit or 8 8-bit MAC's in parallel \citep{tensilica}, thus by pruning in blocks of $B_w=4 \times B_h=1$ at 16-bit or $B_w=8 \times B_h=1$ at 8-bit, we could skip entire cycles of processing.

\subsection{Sparsification Strategies}

For online pruning, we adopt the approach described in \citep{tinylstms}, learning a pruning threshold directly from the data by adding a term to the loss function (Eq. \ref{eq:psa_loss}) which encourages sparsification of the network along with the primary task.

As a baseline for offline pruning we use weight magnitude as an importance criteria where structures (weights, units or blocks) with an L1 norm below a certain threshold are set to zero. The threshold is chosen as a percentile of the distribution of values for each layer of the model. In order to meet the target sparsity for the whole model, we need to find an optimal sparsity per layer that meets the overall target sparsity while achieving the highest performance. This is done by calculating all possible combinations that meet the targeted sparsity with an iterative search going in 10\% steps for all layers while limiting the maximum sparsity allowed per layer to 20\% above the overall target  - for the 80\% and 90\% sparsity targets, a maximum of 95\% per layer is chosen. For weight and block pruning, all layers are considered, while, for unit pruning, the last layer is excluded to maintain the output dimensionality. 

All combinations that meet the target sparsity constraint are evaluated on the test set and the optimal sparsity is chosen based on a heuristic combining STOI \cite{taal2010short}, PESQ \cite{pesq} and SI-SDR \cite{le2019sdr}:
\begin{equation}
    Q = 0.1 \cdot STOI + 0.2 \cdot PESQ + 0.6 \cdot SI{\text -}SDR.
\end{equation}
This is chosen to prioritize separation quality as measured by SI-SDR but also lend some weight to quality (PESQ) and intelligibility (STOI) metrics too.

\begin{table*}
\centering
\resizebox{\linewidth}{!}{%
\begin{tabular}{l c c c c c c c}
\toprule
  & BSS SDR (dB) &  SISDR (dB) & PESQ & STOI &Params (M) &  Size (MB) & Speedup\\
\midrule

Baseline (FP32) \citep{tinylstms}  &   13.70 & - & - & - & 0.97 &  3.70 & 1$\times$  \\

Baseline QEQ (FP32)  &   14.08 & 13.31 & 2.85 & 0.92 & 0.97 &  3.70 & 1$\times$  \\
Baseline QEQ (INT8) &    14.04        &     13.21     & 2.81   & 0.92 &      0.97      &     1.04 & 1$\times$    \\
\midrule
Online Weight Pruned 1 (INT8)   & 14.08 &  13.22 &  2.78 & 0.92 & 0.50 & 0.48 & 0.6$\times$\\
Online Block Pruned 1 (INT8)   & 13.77 & 12.80 &  2.79 & 0.92 & 0.58 & 0.55 & 1.7$\times$\\
Online Unit Pruned 1 (INT8) \citep{tinylstms}  &   13.58  & - & - & - & 0.61 & 0.58 & 2.2$\times$ \\
\midrule
Online Weight Pruned 2 (INT8)   & 13.97 & 13.15 &  2.74 & 0.92 & 0.29 & 0.27 & 0.6$\times$ \\
Online Block Pruned 2 (INT8)   & 13.42 & 12.57 & 2.68 & 0.91 &  0.25  & 0.24 & 2.7$\times$\\
Online Unit Pruned 2 (INT8) \citep{tinylstms} &  13.18 & - & - & - & 0.33 & 0.31 & 3.3$\times$\\
\midrule
Online Weight Pruned 3 (INT8)   & 13.59 & 12.67 &  2.69 & 0.91  & 0.09 & 0.09 & 1.84$\times$ \\
Online Block Pruned 3 (INT8)   & 13.11 & 12.33 & 2.62 & 0.91 &  0.09  & 0.09 & 6.7$\times$\\

\bottomrule

\end{tabular}}
\caption{Online pruning performance comparison on CHiME2 test set. We select the best performing weight and block pruning models that fall under the capacity constraints of the 2 pruned models proposed in TinyLSTMs\cite{tinylstms}.}
\label{table:optimizations}
\end{table*}

\section{Experimental Results}
\label{headings}
Our experiments explore the weight, block, and unit pruning methods using the online and offline sparsification strategies, all using integer quantized models. 
For online pruning, we train each model from scratch with both quantization and pruning objectives in effect.

For objective evaluations, we focus on signal-to-distortion ratio (SDR)\cite{vincent2006performance} to compare model quality across different sparsity percentages, as shown in Fig. \ref{fig:sparsity}. We also include SI-SDR\cite{le2019sdr}, STOI\cite{taal2010short}, and PESQ\cite{pesq} as additional data points in Table \ref{table:optimizations}.

\subsection{Quantization EQ}

The effect of QEQ on the baseline architecture was tested by training the model in both floating point (FP32) and quantized (INT8) arrangements. The floating point model achieved an SDR of 14.08dB SDR, which is a 0.38dB improvement with QEQ compared to previous work using learnable quantization ranges \citep{tinylstms}. In addition, the INT8 model shows very little performance difference across all objective measures compared to the FP32 model.

\subsection{Pruning}

The interplay between sparsity structure, level and  strategy is illustrated in Fig. \ref{fig:sparsity}. We show that our online pruning strategy is able to retain much higher model quality as sparsity increases compared to our offline baseline. The trade-off becomes more significant as the grouping of the weights increases, hence, weight pruning retains the most performance as model size decreases, followed by block pruning, and lastly unit pruning. 

To add additional context to the discussion of sparse structures in online pruning, we include three proposed models from TinyLSTMs \citep{tinylstms} as online unit pruning benchmarks and a baseline.

Offline pruning may be convenient due to its ease and simplicity and can be used to trim up to 25\% from the model's memory footprint post training without much performance loss (0.5 dB SDR) but is strictly inferior to online pruning.

Comparing models with similar performance in Table \ref{table:optimizations}, we note that our \textit{Online Weight Pruned 3} at 13.59 SDR is similar to \textit{Online Unit Pruned 1} at 13.58 dB SDR but with 6.4 $\times$ smaller model size. We also highlight that our \textit{Online Block Pruned 1} actually exceeds the performance of \textit{Baseline} while retaining a 6.2 $\times$ smaller footprint and a speedup of 1.7 $\times$ illustrating the versatility of block pruning.

\section{Conclusions}


Using our methods, we compress an already compact model's memory footprint by a factor of 42$\times$ from 3.7MB to 87kB while   losing negligible performance. We also show a computational speedup of 6.7 $\times$ with a corresponding SDR drop of only 0.59 dB SDR using block pruning. Overall we conclude each pruning method has its own benefits and drawbacks: weight pruning is best for minimizing memory footprint with minimal performance drop, but has a small or negative computational impact. Block pruning strikes a balance between weight and unit pruning, with moderate performance drops at higher pruning ratios but the potential for meaningful throughput increase. And unit pruning provides both throughput increases and memory footprint savings but at a large quality penalty.








\label{headings}

\begin{ack}
This work was partially supported by the Deutsche Forschungsgemeinschaft (DFG, German Research Foundation) under Germany's Excellence Strategy – EXC 2177/1 - Project ID 390895286.

\end{ack}

\bibliography{bib}

\bibliographystyle{unsrtnat} 

\newpage
\section{Appendix}


\begin{figure}[!h]
  \centering
   \begin{subfigure}
         \centering
        \includegraphics[width=\columnwidth]{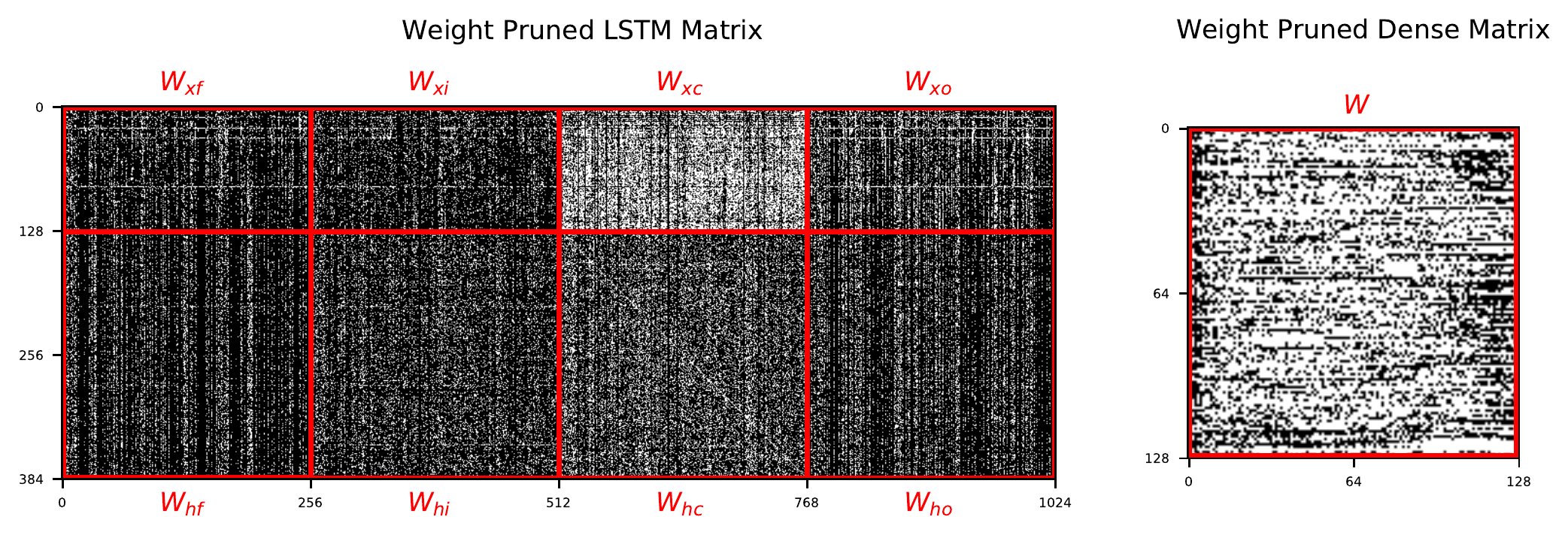}
        \caption{Visualizations of weight sparsity in a \char`~60\% sparse LSTM \textit{(left)} matrix and a \char`~40\% sparse dense \textit{(right)} matrix where black pixels indicate weights that are pruned and white pixels indicate weights that remain. 
        We note that the $W_C$ matrices retain more active nodes than the gate matrices, and the appearance of a distinct diagonal structure in $W_{hc}$. }
        \label{fig:weight}
    \end{subfigure}

    \begin{subfigure}
        \centering
        \includegraphics[width=\columnwidth]{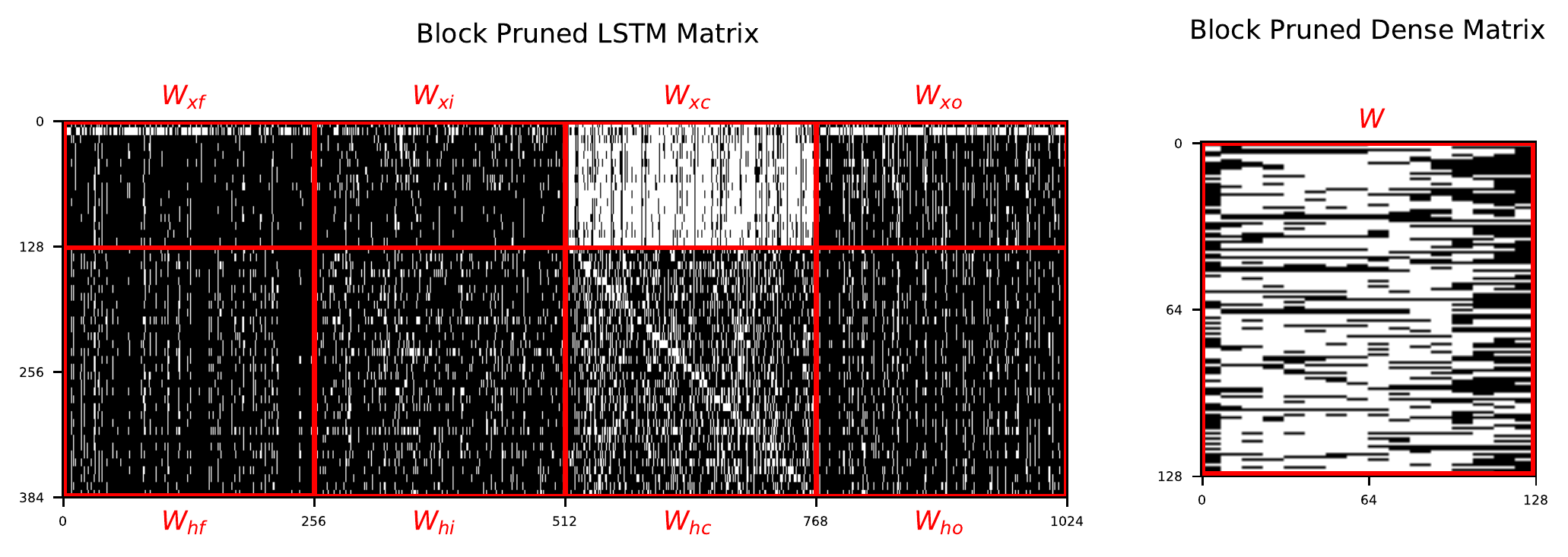}
        \caption{Visualizations of $1 \times 8$ block sparsity in a \char`~60\% sparse LSTM \textit{(left)} matrix and a \char`~40\% sparse dense \textit{(right)} matrix where black pixels indicate weights that are pruned and white pixels indicate weights that remain. 
        We note that the $W_C$ matrices retain more active nodes than the gate matrices, and the appearance of a distinct diagonal structure in $W_{hc}$, even though in this case the diagonal structure contains a coarse approximation of the identity matrix. }
        \label{fig:block}
    \end{subfigure}

    \begin{subfigure}
        \centering
        \includegraphics[width=\columnwidth]{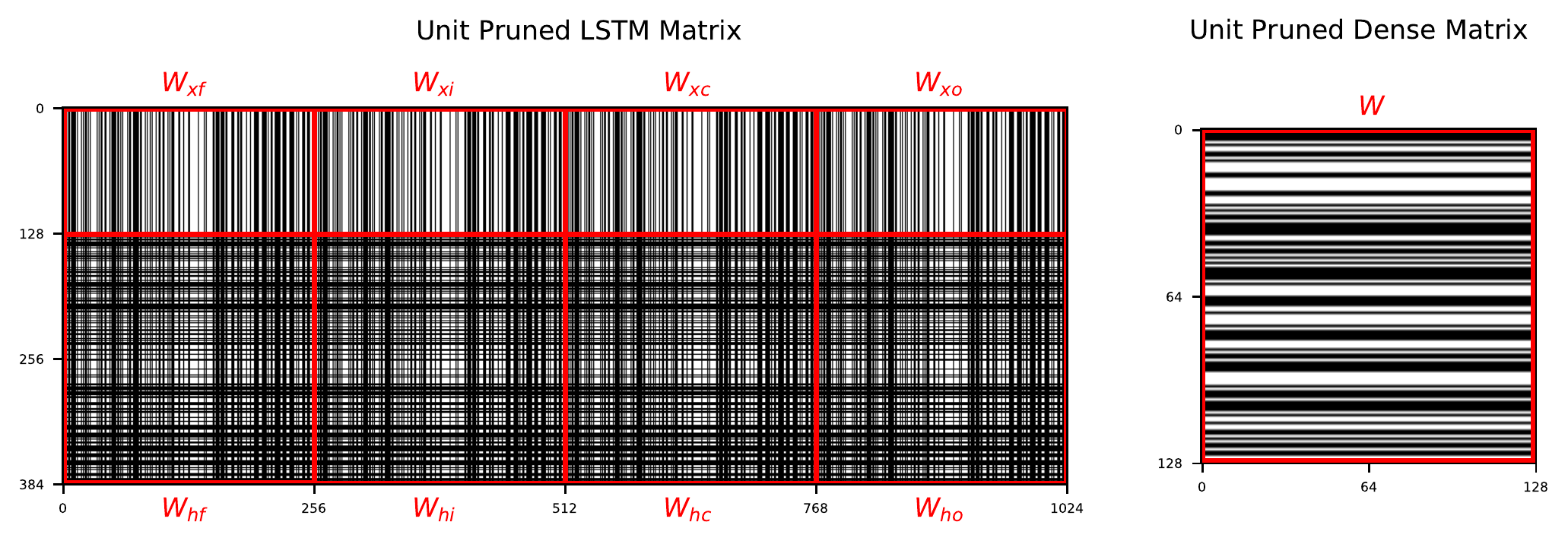}
        \caption{Visualizations of unit sparsity in a \char`~60\% sparse LSTM \textit{(left)} matrix and a \char`~50\% sparse dense \textit{(right)} matrix where black pixels indicate weights that are pruned and white pixels indicate weights that remain. We note the uniformity in pruning across LSTM weight matrices $W_f, W_i, W_C, W_o$ due to the higher level grouping described in Section 3.1.2. We also note the horizontal banding across all $W_h$ matrices, induced by pruned units feeding back into the recurrent part of the cell.}
        \label{fig:row}
    \end{subfigure}
\end{figure}



\end{document}